\documentclass[conference]{IEEEtran}
\IEEEoverridecommandlockouts

\usepackage{cite}
\usepackage{amsmath,amssymb,amsfonts}
\usepackage{algorithmic}
\usepackage{graphicx}
\usepackage{textcomp}
\usepackage{xcolor}
\usepackage{url}
\usepackage{booktabs}
\usepackage{multirow}
\usepackage{pgfplots}
\usepackage{enumitem}
\pgfplotsset{compat=1.18}
\usepackage{pgfplotstable}
\usepackage{subcaption}   
\usepackage{tikz}         

\def\BibTeX{{\rm B\kern-.05em{\sc i\kern-.025em b}\kern-.08em
    T\kern-.1667em\lower.7ex\hbox{E}\kern-.125emX}}

\begin{document}

\title{Can AI Chatbots Provide Coaching in Engineering? Beyond Information Processing Toward Mastery}

\author{
\IEEEauthorblockN{Junaid Qadir\IEEEauthorrefmark{1},
Muhammad Adil Attique\IEEEauthorrefmark{2},
Saleha Shoaib\IEEEauthorrefmark{2},
and Syed Ibrahim Ghaznavi\IEEEauthorrefmark{2}}, 
\IEEEauthorblockA{\IEEEauthorrefmark{1}\textit{Qatar University, Qatar} \\
Email: jqadir@qu.edu.qa}
\IEEEauthorblockA{
\IEEEauthorrefmark{2}\textit{Information Technology University, Pakistan} \\
Email: bscs23196@itu.edu.pk, bscs23036@itu.edu.pk, ibrahim.ghaznavi@itu.edu.pk}
}

\maketitle

\begin{abstract}
\noindent
Engineering education faces a double disruption: traditional apprenticeship models that cultivated judgment and tacit skill are eroding as generative AI emerges as an informal coaching partner. This convergence rekindles long-standing questions in the philosophy of AI and cognition about the limits of computation, the nature of embodied rationality, and the distinction between information processing and wisdom. Building on this rich intellectual tradition, this paper examines whether AI chatbots can provide \textit{coaching} that fosters mastery rather than merely delivering information. We synthesize critical perspectives from decades of scholarship on expertise, tacit knowledge, and human-machine interaction, situating them within the context of contemporary AI-driven education. Empirically, we report findings from a mixed-methods study (\(N=75\) students, \(N=7\) faculty) exploring the use of a coaching chatbot in engineering education. Results reveal a consistent boundary: participants accept AI for technical problem solving (convergent tasks; \(M=3.84\) on a 1--5 Likert scale) but remain skeptical of its capacity for moral, emotional, and contextual judgment (divergent tasks). Faculty express stronger concerns over risk (\(M=4.71\) vs.\ \(M=4.14\), \(p=0.003\)), and privacy emerges as a key requirement, with 64--71\% demanding strict confidentiality. Our findings suggest that while generative AI can democratize access to cognitive and procedural support, it cannot replicate the embodied, value-laden dimensions of human mentorship. We propose a \textit{multiplex coaching} framework that integrates human wisdom within ``expert-in-the-loop'' models---preserving the depth of apprenticeship while leveraging AI's scalability to enrich the next generation of engineering education.\end{abstract}


\section{Introduction}

\subsection{The Changing Landscape of Engineering Formation}

The rapid rise of generative AI (GenAI) is reshaping engineering education in visible and subtle ways \cite{qadir2023engineering}.  Today's engineering students carry artificial intelligence (AI) chatbots in their pockets, experimenting with these tools as informal coaching partners for everything from debugging code to managing academic stress. This informal experimentation is rapidly being institutionalized: platforms like Khanmigo embed AI coaching directly into learning environments \cite{khanacademy2023khanmigo}, while Coursera Coach promises ``personalized support to help you achieve your learning goals" through AI-powered guidance~\cite{coursera2024coach}. Meanwhile, traditional apprenticeship models through which novice engineers once developed judgment are rapidly eroding as automation eliminates entry-level learning opportunities~\cite{beane2023skill}. This double disruption---the erosion of traditional mentorship pathways coinciding with the rapid deployment of AI coaching systems---creates both profound risks and unprecedented possibilities for how we prepare future engineers.

Traditional engineering formation has depended on tacit learning through guided experience---forms of knowing that cannot be fully articulated or automated~\cite{polanyi1966tacit}. Expert engineers who diagnose systems by sound or feel possess knowledge resisting codification~\cite{trevelyan2014making}. Generative AI systems excel at explicit knowledge retrieval and pattern matching but remain limited in cultivating the tacit understanding, ethical discernment, and embodied rationality essential to professional judgment~\cite{dreyfus1986mind}.

\subsection{Risks and Opportunities of AI Coaching}

The stakes are considerable. Over-reliance on AI assistance could accelerate de-skilling, depriving students of productive struggle necessary for mastery and potentially fostering learned helplessness. Yet thoughtfully designed AI coaching systems could democratize personalized support at unprecedented scale, particularly as student mental health concerns intensify and faculty face severe capacity constraints \cite{asghar2024mental}. Contemporary accreditation standards emphasize broad professional competencies beyond technical knowledge---communication, teamwork, ethical judgment, resilience, empathy \cite{abet2019criteria,nae2004engineer}---yet faculty capacity for individualized coaching remains severely constrained. Students are already using AI systems informally; the question is no longer whether AI will play a role, but how we integrate it responsibly.

\subsection{Why Coaching Matters: An Engineering Origin Story}

A striking historical fact illuminates this challenge: modern executive coaching---the practice now being algorithmically replicated---was founded by an engineer. In 1981, Chilean engineer Fernando Flores completed a Berkeley PhD dissertation establishing coaching's intellectual foundations, working with philosopher Hubert Dreyfus to develop ``Conversations for Action" \cite{flores1981}. Flores distinguished between \textit{knowledge as information}---facts and procedures that can be transmitted---and \textit{knowledge as intentionality}---the capacity to transform goals into coordinated action with others \cite{flores1981,Flores2012,jirouskova2023coaching}. This engineering-rooted distinction crystallizes our central question: Can AI chatbots provide coaching that develops intentionality, or are they limited to information delivery?

\subsection{Philosophical Grounding}

This moment revives long-standing debates in the philosophy of AI about the limits of computation, the necessity of embodiment for genuine understanding, and whether wisdom can arise from systems devoid of moral stakes. In \textit{Computer Power and Human Reason}, Weizenbaum~\cite{weizenbaum1976computer} cautioned against the belief that human judgment could be mechanized, arguing that domains grounded in empathy, responsibility, and existential commitment lie beyond computational reach. He emphasized that forms of experience rooted in shared vulnerability and genuine care cannot be reproduced through pattern matching, regardless of algorithmic sophistication. Building on this critique, Winograd and Flores~\cite{winograd1987understanding} showed that human cognition is embedded in embodied action and social practice, not detached information processing. Their work in Understanding Computers and Cognition helped found HCI by demonstrating that humans do not first process information and then act; rather, cognition is inseparable from our ongoing engagement with a meaningful world.

These epistemological limitations extend to coaching's inherently relational dimensions. Effective coaching depends on mutual trust, shared context, and capacity for attunement and care~\cite{schon1983reflective}---precisely the domains where AI faces what recent scholarship identifies as ``in principle obstacles"~\cite{montemayor2022empathic}. Genuine empathy requires emotional resonance, shared vulnerability, and moral responsiveness---qualities computational systems fundamentally lack. When users believe they receive authentic empathy while interacting with affectless simulations, they may over-disclose to systems incapable of appropriate moral response or develop false intimacy mistaking algorithmic responsiveness for human care~\cite{shteynberg2024empathic}. Most troublingly, if AI coaching becomes the default for under-resourced populations while privileged students receive human mentorship, we risk institutionalizing two-tier systems where those who can least afford it are relegated to ersatz alternatives.

Current literature converges on a crucial insight: while AI systems can accelerate convergent reasoning and information retrieval, they remain profoundly deficient in embodied, moral, and relational dimensions underlying professional wisdom~\cite{winograd1987understanding, weizenbaum1976computer}. Effective AI integration requires frameworks that re-anchor technology within human judgment rather than displacing it \cite{qadir2024guides}.

\subsection{Research Questions}

Our investigation focused on these questions:

\begin{enumerate}[label=\textbf{RQ\arabic*:}, leftmargin=1.2cm]

\item \textit{\textbf{Perceptions and Boundaries}.} How do engineering students and faculty perceive AI chatbots' potential role in providing coaching support across different task domains---technical problem-solving, reflective practice, and interpersonal guidance---and where do they identify boundaries and risks? Do these perceptions differ significantly between students and faculty?

\item \textit{\textbf{Acceptance Conditions}.} What concerns about privacy, accountability, and transparency emerge in stakeholder discussions of AI coaching?
\end{enumerate}

\subsection{Multilevel Framework, Contributions, and Paper Overview}

This paper bridges three interdependent inquiry layers. At the \textit{philosophical} level, we revisit enduring questions about computation's limits and embodiment's necessity---drawing on critiques from Weizenbaum~\cite{weizenbaum1976computer}, Winograd and Flores~\cite{winograd1987understanding}, and Dreyfus~\cite{dreyfus1986mind}. At the \textit{theoretical} level, we synthesize these insights with Flores's intentionality framework and Goldberg's Five Shifts model~\cite{goldberg2018,jirouskova2023coaching} to articulate what AI can and cannot do in engineering coaching. At the \textit{empirical} level, we investigate how students and faculty perceive these epistemological boundaries in practice.

This study makes the following \textit{key contributions}:

\begin{enumerate}
    \item It provides the first empirical investigation of AI coaching acceptance in engineering education, revealing a clear trust boundary: participants welcome AI for technical support ($M=3.84$ on a 1--5 Likert scale) but distrust it for moral and emotional guidance ($M_{\text{faculty}}=4.71$ vs.\ $M_{\text{students}}=4.14$, $p=0.003$). 
    
    \item It proposes a \textit{multiplex coaching} model leveraging AI for scalable cognitive support in convergent tasks while preserving human mentorship for divergent developmental guidance involving empathy, ethics, and contextual judgment.
\end{enumerate}

We report findings from a mixed-methods study involving 75 engineering students and 7 faculty members, examining perceptions to inform future evidence-based policy development. This approach prioritizes understanding stakeholder intuitions about epistemological boundaries and ethical concerns before AI coaching tools become institutionally entrenched. 

\textit{Organization}. Section~\ref{sec:background} reviews coaching in engineering education, engineering as embodied practice, and AI in educational contexts. 
Section~\ref{sec:theory} outlines the theoretical framework, defining the convergent--divergent boundary and introducing the multiplex coaching model. 
Section~\ref{sec:methods} details the mixed-methods design, participants, instruments, and analysis. 
Section~\ref{sec:results} presents findings on stakeholder perceptions, boundaries, and safeguards. 
Section~\ref{sec:discussion} discusses implications for responsible AI integration and future research directions. 
Section~\ref{sec:conclusion} concludes with reflections on preserving human judgment and wisdom in partnership with intelligent machines.

\section{Background and Literature Review}
\label{sec:background}

\subsection{Coaching as Developmental Partnership}

Coaching, unlike teaching or tutoring, centers on facilitative guidance that helps learners clarify goals, reflect on experiences, and cultivate self-directed growth~\cite{bachkirova2014complete}. Whereas teaching emphasizes content delivery and tutoring focuses on correcting problems, coaching develops metacognition, motivation, and the capacity to navigate complexity. This is increasingly crucial in outcomes-based engineering education, which demands competencies that extend beyond procedural mastery---creativity, collaboration, ethical reasoning, and adaptive judgment---yet scalable individualized coaching remains difficult to provide~\cite{abet2019criteria,nae2004engineer,qadir2020engineering}.

Flores's seminal distinction between \textit{knowledge as information} and \textit{knowledge as intentionality} clarifies the challenge. Information refers to facts and procedures that can be transmitted; intentionality concerns the capacity to coordinate action, manage commitments, and respond skillfully when plans break down~\cite{Flores2012}. This distinction is central to understanding where AI coaching is inherently limited: while AI excels at information retrieval and pattern generation, intentionality requires embodied experience, social coordination, and accountable action---domains where computational systems lack moral stakes and lived context \cite{qadir2024guides}.

Contemporary coaching theorists reinforce this boundary. Bachkirova and Kemp~\cite{bachkirova2024ai} argue that ``AI coaching'' risks becoming an \textit{ersatz} substitute, replicating the surface structure of coaching while missing its essential relational and ethical foundations. They outline six qualities of genuine coaching---joint inquiry, pragmatic problem-solving, value-based affirmation, contextual insight, trust-based relationship, and mutual accountability. AI systems lack agency, cannot hold values or responsibility, and cannot enter trust-based commitments, placing principled limits on what they can authentically contribute to developmental coaching.

\subsection{Engineering Coaching: Directive and Reflective}

Professional coaching---particularly the Co-Active model~\cite{whitworth2007coactive}---operates through fundamentally non-directive, reflective paradigms emphasizing that clients are naturally creative, resourceful, and whole. Co-Active coaches avoid providing answers, instead asking powerful questions and trusting clients to discover solutions. This approach assumes no objectively ``correct" answers exist---coaching helps individuals align choices with personal values.

Engineering coaching cannot rely solely on such non-directive methods. As Jirouskova and Goldberg~\cite{jirouskova2023coaching} argue, engineering education requires a hybrid approach that integrates reflective, inquiry-based coaching with directive guidance when technical correctness matters. Goldberg's critique in A Whole New Engineer\cite{goldberg2014whole} highlights how engineering education privileges technical rationality at the expense of human capacities like curiosity, listening, and ethical judgment. His later Five Shifts framework\cite{goldberg2018,jirouskova2023coaching} proposes cultivating ``noticing, listening, and questioning'' (NLQ) as core professional capacities for navigating ambiguity, emphasizing attention to language, emotion, and embodied experience.

Yet engineering coaching cannot stop at reflection. Unlike life coaching clients facing predominantly divergent problems with multiple valid paths, engineering students encounter \textit{both} convergent problems (where correct answers exist and must be mastered) and divergent problems (where professional judgment about competing values determines outcomes). Engineering coaches must know when to provide authoritative technical guidance versus reflective techniques, balancing ``wholeness"---engaging students' full humanity---with disciplinary rigor.


\textit{Goldberg's Five Shifts framework}~\cite{goldberg2018,goldberg2023field,jirouskova2023coaching} articulates fundamental transformations required for engineering practice in uncertain environments. Each shift reveals distinct boundaries for AI coaching:

\paragraph{Yogi's Shift (from technical rationality to reflection-in-action~\cite{schon1983reflective})}
Recognizes that professional practice involves thinking and adjusting while doing, not just applying pre-learned formulas. For example, when a design fails, an engineer reflects on what went wrong and why their assumptions broke down. AI can ask structured reflection questions (``What surprised you about this failure?''), but it cannot demonstrate how an experienced practitioner navigates uncertainty in real-time---it has never faced a genuine breakdown requiring on-the-spot adaptation.

\paragraph{Brain-on-a-Stick Shift (from head to heart, body, hands)}
Emphasizes that engineering expertise integrates analytical thinking with emotional awareness, physical intuition, and hands-on experience. Consider an engineer who senses something is ``off'' in a team meeting before anyone explicitly states a problem, or who feels when a machine vibration indicates trouble. AI operates purely through symbol manipulation---it cannot read the physical resistance of materials, sense team tensions, or draw on gut-level intuition that comes from years of hands-on practice.

\paragraph{Wittgenstein's Shift (from language-as-description to language-as-action)}
Drawing on speech act theory, this shift recognizes that professional language doesn't just describe reality---it creates commitments and consequences. When an engineer declares ``I'll have the prototype ready by Friday,'' they stake their reputation and create organizational dependencies. While AI generates linguistically sophisticated responses, its ``commitments'' carry no weight. If it ``promises'' a solution and fails, no one loses their job, no project derails, no trust erodes. Its words lack genuine stakes.

\paragraph{Little Bets Shift (from planning to effectuation)}
In uncertain situations, professionals make small experiments to learn what works, adapting as they go. A student might try three different circuit designs, learning from each failure. AI can analyze experimental data, but it cannot genuinely experiment under uncertainty---it faces no real consequences from failure, the very pressure that drives adaptive learning and builds judgment about when to persist versus pivot.

\paragraph{Polarity Shift (from problem-solving to managing tensions~\cite{johnson1996polarity})}
Many engineering challenges involve ongoing tensions without final solutions---speed versus safety, innovation versus standardization, team autonomy versus coordination. A project manager must continually sense which pole needs emphasis right now based on subtle contextual cues. AI seeks definitive answers and algorithmic resolution, but cannot hold the both-and complexity of knowing when to prioritize efficiency this week but reliability next month based on changing stakeholder dynamics and emerging risks.

This framework suggests AI coaching should focus on \textit{information scaffolding for convergent tasks} while \textit{human coaching develops dispositional shifts} defining engineering judgment.

\subsection{Engineering as Embodied Practice}

A crucial distinction exists between \textit{convergent problems}---having single correct answers achievable through established methods---and \textit{divergent problems}---involving competing values and irreducible trade-offs with no uniquely optimal solution~\cite{johnson1996polarity}. Engineering is fundamentally embodied practice~\cite{polanyi1966tacit,dreyfus1986mind}. Engineers diagnosing failing systems by sound or feel possess tacit knowledge resisting codification, developed through repeated material engagement. Engineering involves social embedding: teamwork, negotiation, organizational dynamics. Expertise develops through pattern recognition grounded in vast experience repertoires, not accumulating explicit rules.


Flores emphasized that simulation and games accelerate mastery by compressing learning time and reducing risk~\cite{jirouskova2023coaching}. Generative AI can create \textit{infinite practice variations} tailored to individual learning edges---generating debugging scenarios, design constraints, or stakeholder dialogues on demand. More powerfully, AI can embody \textit{defined personas}---simulating a demanding client, skeptical regulator, confused team member, environmental advocate---providing safe rehearsal spaces for engineering work's interpersonal dimensions~\cite{trevelyan2014making}.

Consider a student learning to navigate design trade-offs. AI can simulate a project manager prioritizing cost-efficiency, an environmental engineer emphasizing sustainability, and a community representative concerned about local impact---forcing articulation of technical rationales while managing competing stakeholder values. AI cannot replace lived experience with real stakeholders possessing genuine interests and unpredictable complexity, but it can provide \textit{deliberate practice}~\cite{ericsson2006influence} at conversational patterns, reasoning structures, and perspective-taking required for effective engagement.

Similarly, AI can generate realistic failure scenarios: ``Your prototype failed stress testing two weeks before production deadline---identify the root cause and propose a redesign within budget constraints." These simulations build troubleshooting competence through exposure to situations students might encounter once in real practice but can experience dozens of times in simulation. AI can vary parameters systematically in ways human instructors cannot sustain at scale.

Crucially, this positions AI not as autonomous coach but as \textit{coached practice environment}: human instructors design simulation parameters, define learning objectives, and provide debrief protocols where students reflect on performance with guidance. This aligns with Flores's vision of technology as enabler of human coordination rather than replacement for judgment~\cite{winograd1987understanding}.

\subsection{AI in Educational Contexts: Promise and Peril}

AI systems can now simulate aspects of coaching by offering personalized feedback, motivational prompts, and on-demand guidance. For well-structured, convergent tasks, these systems extend access and scalability beyond what human instructors alone can provide. But when AI shifts from tutoring to \textit{coaching}, the risks multiply and require critical scrutiny \cite{holmes2025critical}. Instant, polished responses can bypass the reflective inquiry and productive struggle through which coaching typically develops self-awareness and judgment. Over-reliance also creates Bainbridge’s \textit{ironies of automation} \cite{bainbridge1983ironies}: the more learners outsource planning, interpretation, or decision-making to AI, the less capable they become when faced with novel, ambiguous, or ethically charged situations.

AI coaching tools further embed tacit pedagogical assumptions that privilege explicit, decontextualized knowledge while neglecting the situated, relational, and value-laden dimensions central to genuine coaching \cite{schon1983reflective}. Holmes emphasizes that these technologies cannot be evaluated in isolation \cite{holmes2025critical}: their impact depends on institutional conditions and power structures. If well-resourced institutions use AI to enhance human mentoring while under-resourced institutions substitute AI for human relationships, AI coaching risks entrenching a two-tier system. The key equity question becomes: \textit{Will AI expand access to high-quality developmental support, or will it deliver a ``good enough’’ coaching substitute to those with the fewest resources?} A critical approach must therefore examine not only what AI can do, but whose developmental needs are amplified, whose are diminished, and how authority and agency shift in the process.

\section{Theoretical Framework}
\label{sec:theory}

\subsection{The Convergent-Divergent Epistemological Boundary}

We hypothesize that stakeholder perceptions of AI coaching align with an epistemological boundary between convergent and divergent problem domains~\cite{schumacher1977guide}. Schumacher distinguished between \textit{convergent problems}---having definite solutions where logical analysis drives toward single correct answers (calculating loads, debugging syntax, optimizing algorithms)---and \textit{divergent problems}---involving irreducible contradictions between competing principles requiring ongoing navigation rather than final resolution (balancing innovation against reliability, navigating stakeholder conflicts, determining responsible practice in ambiguous situations).

Brian Cantwell Smith~\cite{smith2019promise} articulates a parallel distinction between \textit{reckoning} (computation, calculation, rule-following) and \textit{judgment} (registration of contextual particularity, embodied understanding of what matters, commitment to consequences). He warns against the ``reckoning fallacy''---believing sufficiently sophisticated computation can substitute for judgment. Engineering practice demands both: reckoning for convergent technical problems and judgment for divergent situations involving values, trade-offs, and human stakes.

AI's strength lies in convergent domains where explicit knowledge is primary, reckoning suffices, and correctness can be verified objectively. AI can rapidly search solution spaces, identify patterns, recall information, and structure guidance according to established procedures---amplifying human capacity without displacing judgment. AI's weakness manifests in divergent domains requiring tacit knowledge, embodied experience, and wisdom to navigate competing goods with no algorithmic resolution~\cite{dreyfus1986mind, polanyi1966tacit}.

Schumacher and Johnson~\cite{schumacher1977guide,johnson1996polarity} identify a fundamental error: treating divergent problems as if they were convergent---attempting to ``solve'' through logic what requires ongoing navigation, or seeking final resolution to irreducible polarities that must be continually balanced. Johnson demonstrates that many engineering challenges involve tensions between interdependent opposites (efficiency and thoroughness, innovation and standardization, speed and quality) where optimizing one pole inevitably creates problems requiring the other. AI systems, designed for problem-solving, cannot hold such polarities in productive tension. In divergent domains, AI thus risks creating false confidence by offering definitive answers to questions admitting no final resolution, or displacing the human relationships through which practitioners develop wisdom to navigate enduring tensions.

Our framework predicts stakeholders will intuitively recognize this boundary. We expect acceptance of AI for technical problem-solving, concept clarification, and structured reflection on well-defined tasks, but resistance in domains requiring empathy, moral discernment, and interpersonal navigation---precisely where judgment is essential and coaching, as distinct from tutoring, is most needed.

\subsection{A Multiplex Coaching Framework}

Based on our theoretical synthesis, we propose a hybrid model that strategically divides labor between AI systems and human mentors according to their respective capabilities and limitations. This is not a model of AI as replacement but of AI as carefully circumscribed support within an architecture that preserves human judgment as primary.

We hypothesize that AI can support \textit{convergent cognitive scaffolding}: providing on-demand technical Q\&A for concept clarification and formula application, offering structured reflective prompts after assignments to guide metacognitive processing, delivering immediate feedback on well-defined problems where correctness is algorithmically verifiable, sending accountability reminders for deadlines and commitments, assisting with brainstorming for idea generation within established frameworks, and identifying patterns across problem sets to support transfer. These functions would leverage AI's computational strengths---tireless availability, instant processing, and consistent application of defined criteria---while remaining within domains where verification against objective standards remains possible.

Human mentors would retain primary responsibility for \textit{divergent developmental guidance} requiring embodied judgment and moral responsiveness. This includes emotional support requiring genuine empathy~\cite{montemayor2022empathic,shteynberg2024empathic}, career guidance and professional identity formation, ethical dilemmas involving competing values and stakeholder trade-offs, interpersonal conflict navigation, team dynamics, mental health concerns requiring appropriate referral to specialized support, existential questions about meaning and purpose, and judgment calls in high-stakes ambiguous situations where context matters. These domains resist algorithmic resolution not due to current technical limitations but because they fundamentally require the intentionality, embodied presence, and moral commitment that characterize genuine human relationships~\cite{winograd1987understanding,weizenbaum1976computer}.

The integration principle positions AI as supplement rather than substitute, with AI potentially handling scalable cognitive support for convergent tasks while freeing faculty for high-touch developmental guidance where human presence may prove essential. Some tasks will inevitably fall in gray zones requiring careful judgment---post-failure reflection might begin with AI-guided prompts but require human follow-up when emotional distress or persistent misconceptions emerge. The framework functions as heuristic for institutional decision-making rather than prescriptive algorithm, with appropriate boundaries between AI and human support remaining an empirical question requiring systematic investigation of stakeholder perspectives and developmental outcomes.

\subsection{AI Literacy as Moderating Factor}

We expect that AI literacy---understanding of both capabilities and limitations---will moderate stakeholder perceptions. Informed users who have extensive experience with AI systems may develop more accurate and critical assessments, recognizing both genuine utility and real constraints \cite{qadir2024learning,tabassum2024critical}.

Paradoxically, we hypothesize that experienced AI users may be \textit{more} skeptical of AI coaching in high-stakes divergent domains, having witnessed failures firsthand. Those who understand how large language models (LLMs) work (as statistical pattern matchers trained on text rather than as reasoning agents with genuine understanding) may be particularly cautious about relying on AI for moral guidance or complex judgment calls \cite{bender2021dangers}.

This suggests that AI literacy education should be dual-faceted: helping students learn both how to use AI tools effectively \textit{and} how to recognize when these tools should not be used. The goal is not to create either uncritical enthusiasm or blanket rejection but to cultivate discernment about where AI support is appropriate and where human judgment remains essential \cite{qadir2024educating,holmes2025critical}.

\section{Methodology}
\label{sec:methods}

\subsection{Research Design and Rationale}

We conducted a mixed-methods perception study to assess stakeholder attitudes toward AI coaching in engineering education prior to formal institutional integration. This approach prioritizes understanding demand, mapping perceived boundaries, and identifying ethical concerns before institutionalization. Our design combines quantitative survey measures (Likert scales) to assess perceptions across dimensions aligned with our theoretical framework, with brief qualitative open-ended responses to capture nuanced concerns and boundary conditions. This mixed approach allows us to identify statistical patterns while preserving contextual richness that purely quantitative methods might obscure. The study is fundamentally exploratory, aiming to generate empirically-grounded hypotheses about appropriate human-AI boundaries rather than testing causal claims about coaching effectiveness. Given that many students are already using AI tools informally---often without guidance or institutional oversight---systematic investigation of stakeholder perceptions is essential for responsible integration and provides empirical grounding for the multiplex coaching framework proposed above. 

\subsection{Participants and Sampling}

\textit{Student Sample:} We recruited 75 engineering students from a university engineering program at Information Technology University, Pakistan, representing multiple years of study (first-year through graduate level) and diverse exposure to AI tools. Participants were recruited through course announcements and received no compensation, ensuring voluntary participation motivated by interest in the topic rather than external incentives. The student sample ($N=75$) provides sufficient size for detecting meaningful patterns in perceptions, though more subtle differences may be difficult to discern. Diversity in year of study allows us to explore whether perceptions evolve as students progress through their programs and gain greater engineering maturity.

\textit{Faculty Sample:} Seven engineering instructors with varying teaching experience and diverse levels of AI tool adoption participated in parallel surveys. This small faculty sample limits statistical power and generalizability, but provides valuable preliminary insights into instructor perspectives, which may differ systematically from those of students.

\textit{Sampling Limitations:} Both samples are convenience samples from a single institution, limiting generalizability to other contexts. The voluntary nature of participation may introduce self-selection bias, with those already interested in AI being overrepresented. These limitations are acknowledged, with findings treated as suggestive rather than definitive. Importantly, this study examines stakeholder perceptions and acceptance rather than measuring actual learning outcomes or coaching effectiveness.

\subsection{Instrument Development and Constructs}

We developed structured survey instruments assessing four primary constructs aligned with our theoretical framework:

\textit{Perceived Usefulness Across Task Domains:} Items assessed perceived utility of AI coaching for convergent tasks (technical problem-solving, concept clarification, structured post-failure reflection) versus divergent tasks (managing emotional frustration, navigating team conflicts, reflecting on learning habits and motivation). This construct explores stakeholder perceptions of the convergent-divergent theoretical boundary.

\textit{Trust and Relational Boundaries:} Items probed comfort with different types of disclosure (academic struggles versus personal challenges), trust in AI advice across domains, and perceived need for human mentorship alongside or instead of AI support. This construct captures stakeholder intuitions about where relational authenticity matters.

\textit{Risk Perception:} Items assessed concerns about over-reliance and dependency, accuracy and hallucination risks, privacy and surveillance, and reduction in human interaction quality. This construct identifies perceived threats to student wellbeing and educational integrity.

\textit{Institutional Safeguards:} Items measured demanded privacy protections, transparency expectations, and acceptable governance models. This construct informs policy design for responsible deployment.

All items used 5-point Likert scales (1=Strongly Disagree to 5=Strongly Agree), providing ordinal data suitable for non-parametric analysis while allowing calculation of means for interpretability. Items were adapted from established technology acceptance literature and pilot-tested with a small sample to ensure clarity and comprehension.

\textit{Composite Scale Construction:} To reduce complexity and improve reliability, we constructed two composite indices. \textit{Composite Utility} averaged all perceived usefulness items, providing an overall measure of AI coaching acceptance. \textit{Composite Risk} averaged all risk perception items, capturing overall concern about potential harms. Internal consistency for both composites was assessed using Cronbach's alpha.

\subsection{Data Analysis Strategy}

Our mixed-methods analysis integrated quantitative and qualitative approaches:

\subsubsection{Descriptive and Reliability Analysis} We calculated means, standard deviations, and frequency distributions for all items, disaggregated by cohort (students versus faculty) and by year of study within the student sample. This provided initial portraits of stakeholder perceptions. Cronbach's alpha coefficients were computed for composite scales to assess internal consistency. Acceptable reliability ($\alpha > 0.70$) supports treating composites as coherent constructs.

\subsubsection{Inferential Statistics} We employed Welch's t-tests to compare students and faculty on composite scores (Utility and Risk), using the Welch variant to accommodate unequal variances between groups of different sizes. This addresses RQ2 regarding systematic stakeholder differences. Statistical significance was assessed at $\alpha = 0.05$, with effect sizes (Cohen's $d$) reported to contextualize practical importance alongside statistical significance.

\subsubsection{Qualitative Analysis} Brief open-ended responses were thematically coded to identify recurring concerns, boundary conditions, and illustrative examples. These provide interpretive depth that helps explain stakeholder reasoning behind quantitative patterns and captures their perspectives in their own terms.

\section{Results}
\label{sec:results}

\subsection{Composite Scale Reliability and Group Comparisons}

This section examines the internal consistency of composite scales and compares student and faculty perceptions of AI coaching utility and risk. Table~\ref{tab:composite_analysis} summarizes descriptive statistics, reliability coefficients, and inferential results for both composite scales.



\begin{table*}[t]
\caption{Composite Scale Analysis: Descriptive Statistics, Reliability, and Group Comparisons}
\label{tab:composite_analysis}
\centering
\small
\setlength{\tabcolsep}{8pt}
\begin{tabular}{@{}lcccccc@{}}
\toprule
\textbf{Scale} & \textbf{Group} & \textbf{M (SD)} & \textbf{Cronbach's $\alpha$} & \textbf{t} & \textbf{p} & \textbf{Cohen's d} \\
\midrule
\multirow{2}{*}{\textbf{Composite Utilization}}
& Student & 3.84 (0.62) & 0.82 & \multirow{2}{*}{ns} & \multirow{2}{*}{ns} & \multirow{2}{*}{ns} \\
& Faculty & 3.83 (0.58) & 0.79 & & & \\
\midrule
\multirow{2}{*}{\textbf{Composite Perceived Risk}}
& Student & 4.14 (0.54) & 0.76 & \multirow{2}{*}{$-3.98$} & \multirow{2}{*}{0.003**} & \multirow{2}{*}{1.34} \\
& Faculty & 4.71 (0.28) & ---\textsuperscript{\textdagger} & & & \\
\bottomrule
\end{tabular}
\newline
\vspace{1mm}
\footnotesize
\textit{Note:} **$p < 0.01$; ns = not significant; Between-group comparison statistics ($t$, $p$, Cohen's $d$) based on Welch's $t$-test. \newline \textsuperscript{\textdagger}Faculty Risk Cronbach's $\alpha$ not reported due to small sample size ($N = 7$).
\end{table*}

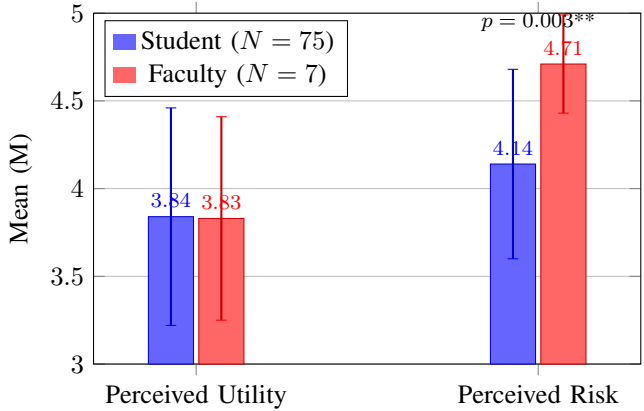
\begin{figure}[t]
\centering
\begin{tikzpicture}
\begin{axis}[
    ybar,
    bar width=0.6cm,
    width=\linewidth,
    height=6.25cm,
    ylabel={Mean (M)},
    ymin=3, ymax=5,
    ymajorgrids=true,
    xtick=data,
    symbolic x coords={Perceived Utility, Perceived Risk},
    xticklabel style={align=center, font=\normalsize},
    enlarge x limits=0.3,
    legend style={at={(0.02,0.98)}, anchor=north west, legend columns=1},
    legend image code/.code={
        \draw[draw=none] (0cm,-0.1cm) rectangle (0.3cm,0.1cm);
    },
    nodes near coords,
    nodes near coords style={font=\footnotesize, anchor=south},
    every node near coord/.append style={/pgf/number format/fixed, /pgf/number format/precision=2}
]

\addplot+[
    fill=blue!60, 
    draw=blue!80!black,
    error bars/.cd,
    y dir=both, 
    y explicit,
    error bar style={line width=0.8pt}
] coordinates {
    (Perceived Utility, 3.84) +- (0, 0.62)
    (Perceived Risk, 4.14) +- (0, 0.54)
};
\addlegendentry{Student ($N=75$)}

\addplot+[
    fill=red!60, 
    draw=red!80!black,
    error bars/.cd,
    y dir=both, 
    y explicit,
    error bar style={line width=0.8pt}
] coordinates {
    (Perceived Utility, 3.83) +- (0, 0.58)
    (Perceived Risk, 4.71) +- (0, 0.28)
};
\addlegendentry{Faculty ($N=7$)}

\node[font=\footnotesize] at (axis cs:Perceived Risk, 4.95) {$p=0.003$**};

\end{axis}
\end{tikzpicture}
\caption{\textbf{Perceived utility and risk of AI coaching} \textit{among engineering students and faculty}. Error bars represent standard deviation (SD). While both groups find AI coaching similarly useful (no significant difference, $p=0.985$), faculty perceive markedly greater risks than students ($t=-3.98$, $p=0.003$, Cohen's $d=1.34$), revealing a clear trust boundary. **$p<0.01$.}
\label{fig:ai_utility_risk}
\end{figure}

\begin{figure}[t]
\centering
\begin{tikzpicture}
\begin{axis}[
    ybar,
    bar width=0.6cm,
    width=\linewidth,
    height=6cm,
    ylabel={Percentage of Cohort (\%)},
    ymin=0, ymax=100,
    ymajorgrids=true,
    xtick=data,
    symbolic x coords={Must remain private, Anonymized data OK, Sharing if improves, Full instructor access},
    xticklabel style={align=center, font=\small, rotate=0},
    x tick label style={text width=2.5cm, align=center},
    enlarge x limits=0.15,
    legend style={at={(0.98,0.98)}, anchor=north east, legend columns=1},
    legend image code/.code={
        \draw[draw=none] (0cm,-0.1cm) rectangle (0.3cm,0.1cm);
    },
    nodes near coords,
    nodes near coords style={font=\footnotesize},
    every node near coord/.append style={/pgf/number format/fixed, /pgf/number format/precision=1}
]

\addplot[fill=blue!60, draw=blue!80!black] coordinates {
    (Must remain private, 64.6)
    (Anonymized data OK, 29.2)
    (Sharing if improves, 27.1)
    (Full instructor access, 16.7)
};
\addlegendentry{Student ($N=75$)}

\addplot[fill=red!60, draw=red!80!black] coordinates {
    (Must remain private, 71.4)
    (Anonymized data OK, 42.9)
    (Sharing if improves, 0.0)
    (Full instructor access, 28.6)
};
\addlegendentry{Faculty ($N=7$)}

\end{axis}
\end{tikzpicture}
\caption{\textbf{Views on Privacy and Data Governance:} \textit{Faculty vs. Student}. Frequency distribution of privacy preferences regarding AI coaching conversation logs (multi-select responses). The results confirm that both cohorts prioritize psychological safety and confidentiality, with 71.4\% of faculty and 64.6\% of students demanding that conversations ``Must remain completely private and not accessible to instructors.'' This strong consensus serves as the primary justification for institutional data governance and accountability policies.}
\label{fig:privacy_preferences}
\end{figure}
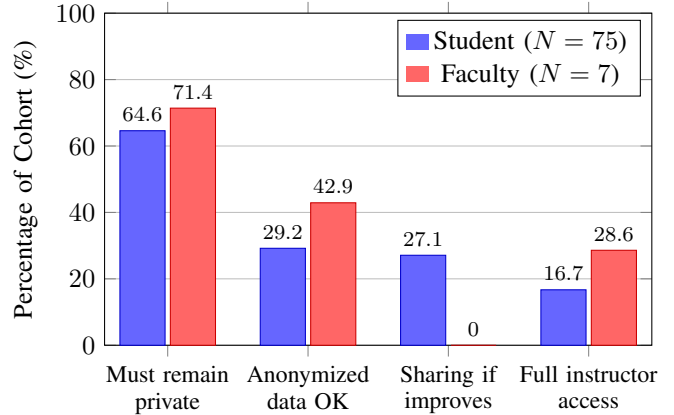

\subsubsection{Reliability Analysis}

Internal consistency was acceptable for both Composite Utility scales ($\alpha=0.82$ for students, $\alpha=0.79$ for faculty) and the student Composite Risk scale ($\alpha=0.76$), supporting their use as coherent constructs. The faculty Composite Risk scale could not be reliably assessed ($N=7$), as small samples produce unstable alpha estimates. However, the descriptive pattern remains informative: faculty expressed uniformly strong concerns across risk items, resulting in minimal variance ($SD=0.28$) compared to students ($SD=0.54$).

\subsubsection{Student-Faculty Comparisons}

Welch's $t$-tests compared student and faculty composite scores, accounting for unequal sample sizes and heterogeneous variances. Two distinct patterns emerged: 

(1) \textit{Utility Consensus:} No significant difference was observed between student ($M=3.84$, $SD=0.62$) and faculty ($M=3.83$, $SD=0.58$) perceptions of AI coaching utility, $t(9.85)=0.02$, $p=0.985$, Cohen's $d=0.02$. This negligible effect size indicates stakeholder consensus: both groups perceive similar functional value in AI for technical and academic support tasks. 

\textit{Risk Divergence:} Faculty expressed significantly higher risk concern ($M=4.71$, $SD=0.28$) than students ($M=4.14$, $SD=0.54$), $t(7.65)=-3.98$, $p=0.003$, Cohen's $d=1.34$. This large effect size indicates substantial divergence in risk perception. Faculty risk perceptions approach ceiling effects ($M=4.71$ on a 5-point scale), suggesting near-universal concern about potential harms---concerns that students acknowledge but perceive as less severe. This divergence has institutional implications: faculty skepticism about developmental risks warrants serious consideration in deployment policy. Fig. \ref{fig:ai_utility_risk} illustrates a clear asymmetry between perceived utility and perceived risk of AI coaching. Students and faculty evaluated AI’s utility at nearly identical levels, whereas faculty perceived significantly higher risks than students, indicating a pronounced trust boundary around AI’s role in coaching.

\subsubsection{Limitations}

The small faculty sample ($N=7$) limits statistical power and generalizability. The significant risk difference, despite low power, reflects the effect's large magnitude and low faculty variance. Future studies should replicate these findings with larger and more diverse faculty samples to assess generalizability beyond this single-institution context.

\subsection{Item-Level Perceptions}

\subsubsection{\textbf{Student Perceptions}} 
Students showed high acceptance for convergent support tasks: getting unstuck on technical problems ($M>4.0$), analyzing failure and planning improvement ($M>3.8$), and concept clarification ($M>4.0$). Moderate acceptance appeared for structured reflection on learning habits ($M\approx 3.5$). Low trust characterized divergent domains: team conflict management ($M=3.10$, $SD=1.08$) indicated a boundary where high-context challenges require human judgment. Critically, students strongly desired human mentors even with AI availability ($M>4.0$), suggesting they view AI as supplement rather than replacement.

\subsubsection{\textbf{Faculty Perceptions}} 
Faculty agreed AI could reduce workload through repetitive clarification questions ($M>4.0$) and support reflection after failure ($M>3.5$). However, a stark boundary emerged: 0\% of faculty trusted AI for emotional support roles, establishing the affective domain as requiring human presence. Faculty showed high concern about student over-trust (``Students might over-trust AI even when wrong'' received strong agreement) and universal consensus that ``AI should complement, not replace, human mentorship.''

\textit{Qualitative Insights on AI Coaching Boundaries.} To complement our quantitative findings, we included open-ended questions in our faculty survey at ITU exploring appropriate boundaries for AI intervention in student learning. Responses reinforced the convergent-divergent distinction emerging from our statistical analysis while revealing nuanced boundary conditions. 

When asked \textit{\underline{``\textbf{where AI could realistically help students}?}''}, faculty identified convergent technical tasks: \textit{``Explaining complex topics, paper preparation by generating questions, understanding subtle issues, explore better solutions available after they are done solving a problem''} and \textit{``Learning concepts, improving writing skills, practicing technical discussions.''} One faculty member emphasized conditionality: \textit{``if AI is tailored according to specific requirements mainly meant to be used for [...] Programming and Maths.''} These responses align with our quantitative findings showing faculty acceptance of AI for technical skill development and structured problem-solving.

When asked \textit{\underline{``\textbf{where AI should never intervene}?}''}, faculty explicitly rejected AI for divergent challenges requiring human judgment: \textit{``Emotional health, complete assignment solving''} and \textit{``Should not be used to solve assignments and problems that students are expected to solve independently.''} One faculty member articulated the cognitive development rationale: \textit{``Ideas Thinking, Debugging Programming Errors. This is due [sic] the fact that problem solving will always be a human-level cognitive learning and will never be replaced by AI.''} Another emphasized relational boundaries: \textit{``should not be a replacement for book reading and an instructor, shouldn't intervene in emotional matters.''}

\subsection{Interpreting Stakeholder Perceptions}

\subsubsection{Conditional Acceptance: The Convergent--Divergent Boundary}

Findings suggest that ``AI coaching'' is inherently conditional rather than universal~\cite{bachkirova2024ai}. High utility ratings (M$\approx$3.84) indicate AI is perceived as effective for convergent, well-structured tasks, whereas low trust for divergent tasks underscores its perceived inadequacy as a genuine \textit{coach}. Participants clearly distinguished between acceptable uses (e.g., ``getting unstuck''---information processing) and unacceptable ones (e.g., ``guiding values''---judgment and empathy). This boundary appears consistent with Schumacher's framework: students welcome AI for solvable problems but reject it for moral or contextual dilemmas, a distinction even more pronounced among faculty. These perceptions suggest that AI coaching design should respect this epistemological limit.

\subsubsection{Acceptable Use: Structured Reflection}

Moderate acceptance for post-failure reflection ($M>3.5$) indicates a potential niche for AI in prompting metacognition---helping students analyze mistakes, generate strategies, and plan improvements. This aligns with mastery learning principles~\cite{bandura1997self}, where reflection transforms error into insight. Yet AI's role appears procedural, not empathic: it can initiate reflection, but it cannot \textit{feel} or authentically respond to a learner's experience.

\subsubsection{Perceived Boundaries: Empathy and Judgment}

The \textit{emotional red line}---a complete lack of faculty trust for emotional support---aligns with Halpern's~\cite{halpern2024empathy} argument that empathy cannot be mechanized. AI lacks shared vulnerability, moral responsibility, and genuine care. Students' preference for human mentors ($M>4.0$) suggests that emotional and ethical growth remain intrinsically human. Similarly, low trust for AI in managing team conflict ($M=3.10$) highlights the perceived need for lived experience and tacit understanding, echoing Trevelyan's~\cite{trevelyan2014making} notion of embodied, situated knowledge that defies codification.

\subsubsection{Recognized Risks: De-skilling Concerns}

Students openly recognize the danger of over-reliance (Composite Risk $M=4.14$), echoing Bainbridge's~\cite{bainbridge1983ironies} and Beane's~\cite{beane2023skill} warnings that automation erodes expertise when it replaces human practice. To counter this, AI tools should foster \textit{productive struggle} through Socratic questioning (e.g., ``What's your hypothesis? Let's test it step-by-step'') rather than delivering direct answers. Such interaction cultivates verification, critical thinking, and AI literacy. Faculty's higher risk perception ($M=4.71$) reflects their responsibility to maintain academic rigor, reinforcing the need for institutional policies that define clear limits for AI's pedagogical use.

\subsubsection{Interpreting Perceptions Through the 5 Shifts Framework}

Our findings align largely with Goldberg's Five Shifts framework~\cite{goldberg2018,jirouskova2023coaching}. The \textit{Yogi's Shift}---from technical rationality to reflection-in-action---explains why participants accepted AI for technical problem-solving ($M=3.84$) but showed moderate acceptance for reflection: AI prompts reflection but cannot model improvisational adaptation. Rejection of AI for emotional and ethical domains corresponds to the \textit{Brain-on-a-Stick} and \textit{Polarity Shifts}---participants' skepticism reflects recognition that integrating head-heart-body and managing irreducible value tensions requires embodied judgment computational systems lack. The faculty-student risk perception difference ($M=4.71$ vs. $M=4.14$, $p=0.003$) gains explanation here: faculty who underwent these dispositional shifts may possess refined awareness of what transformations require, heightening sensitivity to AI's limitations. This consonance demonstrates the Five Shifts framework's utility for conceptualizing AI's appropriate role and provides preliminary empirical support.

\subsection{Privacy and Accountability}

Strong demand for complete privacy emerged as a precondition for authentic disclosure: 71.4\% of faculty and 64.6\% of students indicated conversations must remain completely private from instructors. This finding creates a design constraint---psychological safety requires robust data governance addressing surveillance concerns before deployment. Students are unlikely to seek help if monitored, yet institutions require oversight to ensure accountability. Balancing these priorities calls for transparent governance mechanisms, including:  
\textit{(a)} anonymous data logging without identifiable records;  
\textit{(b)} opt-in participation and explicit consent;  
\textit{(c)} strict data retention and deletion policies;  
\textit{(d)} separation of AI use from assessment; and  
\textit{(e)} clear disclaimers emphasizing human verification.  
Faculty must also be trained in AI's limits and equipped with escalation pathways to human counselors when necessary.

\section{Discussion}
\label{sec:discussion}

\subsection{Parallels from LLM-Based Health Coaching}

Recent work in LLM-based health coaching offers instructive parallels for engineering education. Jörke et al.\cite{jorke2025gptcoach} identified three coaching roles---facilitator, educator, and supporter---that align with Goldberg's hybrid model and demonstrated LLMs' capacity to gather rich qualitative context beyond performance metrics. However, a four-week randomized study ($N=54$) found that while LLM-augmented coaching improved psychological outcomes (stronger beliefs about benefits, greater enjoyment, increased self-compassion), it produced no significant behavioral differences compared to non-LLM controls, with both conditions equally doubling participants meeting activity guidelines\cite{jorke2025bloom}. This psychological-behavioral divergence appears consistent with our finding that students accept AI for technical support while rejecting it for emotional and ethical guidance. The health coaching literature thus suggests effective LLM augmentation requires validated frameworks (e.g., Motivational Interviewing), multimodal interactions beyond text, and recognition that engagement does not guarantee outcomes---considerations essential for multiplex coaching models allocating AI and human expertise by problem type rather than technological capability~\cite{jorke2025gptcoach,jorke2025bloom}.

\subsection{Implications: Designing Multiplex Coaching Systems}

The convergent-divergent distinction suggests the value of \textit{multiplex coaching}: strategic allocation of AI and human expertise based on problem type. Jirouskova and Goldberg~\cite{jirouskova2023coaching} argue engineering coaching requires hybrid practice spanning directive and reflective modes, which can be operationalized through multiplex ``both-and'' thinking~\cite{qadir2024educating}. AI could support convergent problem-solving through immediate technical feedback, infinite practice scenarios, and structured reflection prompts, while humans remain essential for divergent problems requiring situated judgment, what Goldberg terms embodied ``listening''~\cite{goldberg2014whole} that reads emotional cues signaling deeper uncertainties, ethical reasoning grounded in moral accountability, and apprenticeship-like modeling of navigating genuine uncertainty~\cite{montemayor2022empathic,shteynberg2024empathic}. Effective implementation would require transition mechanisms escalating beyond AI's competence (``This involves competing values; consult your instructor'') and cultivating \textit{critical AI literacy}~\cite{qadir2024educating}---understanding AI outputs as ``useful not true'' provocations while maintaining productive struggle rather than outsourcing cognitive effort.

\subsection{Limitations}

Our study had a number of limitations. First, the study assessed perceptions rather than actual behaviors; observed use may differ in practice. Second, the small faculty sample ($N=7$) limits generalizability. Third, its single-institution, cross-sectional scope restricts temporal and contextual reach. Finally, brief open-ended responses constrained qualitative depth; future studies should include interviews or focus groups for richer insight.

\subsection{Future Research}
Future research should extend these findings by implementing a prototype AI coaching system in authentic learning environments to observe real behavioral use, learning outcomes, and well-being effects over time. This phase would move beyond perception-based analysis toward empirical validation, expanding faculty participation and employing longitudinal, multi-institutional designs for broader generalizability. Incorporating the identified practices---\textit{privacy-first data governance}, \textit{transparent feedback loops}, and \textit{Socratic prompting}---could test how well these safeguards and pedagogical principles alleviate the limitations of current chatbots. Ultimately, such work can inform the development of next-generation coaching systems that complement, rather than compete with, human mentorship, nurturing both technical competence and moral character in engineering education.

\section{Conclusions}
\label{sec:conclusion}

Engineering education faces a double disruption: eroding apprenticeship pathways, coinciding with students' unstructured use of AI tools. This study provides evidence-based guidance, offering a nuanced answer to whether AI chatbots can genuinely coach engineering students. We find a conditional yes for convergent support---technical troubleshooting, concept clarification, structured reflection---but a clear no for divergent, judgment-laden guidance involving emotional support, ethical dilemmas, and interpersonal conflict. AI extends access to certain coaching functions at scale but cannot substitute for human guidance in cultivating professional judgment, practical wisdom, and resilience. AI coaching represents strategic augmentation rather than replacement, enabling hybrid models that preserve human mentorship for complex developmental guidance: professional identity formation, ethical reasoning, and sense of belonging. Success requires transparency about AI's limitations, privacy protections, faculty training, and ongoing evaluation. Engineering educators should neither fear nor naively embrace AI coaching, but reimagine support structures---using AI to democratize academic help while elevating human mentorship for irreplaceable dimensions of engineering formation. Engineering mastery requires more than information processing; judgment, creativity, empathy, and wisdom develop through human connection. AI can extend our reach and free faculty for high-touch guidance, but cannot replace embodied presence. The path forward lies in multiplex models that strategically leverage both computational and human capacities while honoring the irreplaceable human element in cultivating capable, ethically responsible engineers.

\section*{Acknowledgment}
The authors thank the participating faculty and students and acknowledge the use of AI tools (ChatGPT, Claude, Grammarly) for writing, editing, and formatting support. All the content has been verified, and the authors take full responsibility for accuracy and integrity.

\bibliographystyle{IEEEtran}
\bibliography{main}

\end{document}